\documentclass[twocolumn,superscriptaddress,longbibliography,aps,pra,preprintnumbers]{revtex4-2}

\usepackage[normalem]{ulem}
\usepackage{graphicx}
\usepackage{bm}
\usepackage{color}
\usepackage{epstopdf}
\usepackage{amsmath}
\usepackage{amssymb}
\usepackage{epstopdf}
\usepackage{lipsum}

\usepackage{multirow}

\usepackage[urlcolor=blue,colorlinks=true,citecolor=blue,linkcolor=blue,pdfstartview={FitH},bookmarks=false]{hyperref}

\usepackage{xcolor}

\graphicspath{{fig/}{./fig/}{.}}

\begin{document}

\title{
Dynamical properties of magnetic topological insulator $T$Bi$_{2}$Te$_{4}$ ($T=$Mn, Fe): \\
phonons dispersion, Raman active modes, and chiral phonons study
}

\author{Aksel Kobia\l{}ka}
\email[e-mail: ]{akob@kft.umcs.lublin.pl}
\affiliation{Institute of Physics, Maria Curie-Sk\l{}odowska University,
Plac Marii Sk\l{}odowskiej-Curie 1, PL-20031 Lublin, Poland}

\author{Ma\l{}gorzata~Sternik}
\email[e-mail: ]{sternik@wolf.ifj.edu.pl}
\affiliation{Institute of Nuclear Physics, Polish Academy of Sciences, W. E. Radzikowskiego 152, PL-31342 Krak\'{o}w, Poland}

\author{Andrzej Ptok$\,$}
\email[e-mail: ]{aptok@mmj.pl}
\affiliation{Institute of Nuclear Physics, Polish Academy of Sciences, W. E. Radzikowskiego 152, PL-31342 Krak\'{o}w, Poland}

\date{\today}

\begin{abstract}
Recently discovered magnetic topological insulators $T$Bi$_{2}$Te$_{4}$ ($T=$Mn, Fe) crystallize into the $R\bar{3}m$ rhombohedral structure and exhibit the antiferromagnetic order.
Here, we discuss the lattice dynamics of these compounds to confirm the stability of these systems. 
We show that the phonon dispersion does not contain soft modes, so both compounds are dynamically stable in the $R\bar{3}m$ phase.
We perform theoretical analyses of the mode activity at $\Gamma$ point for the discussed compounds.
In the case of the Raman active modes, our results are in agreement with the experimentally observed frequencies.
Finally, we also discuss the possibility of realization of chiral phonons.
\end{abstract}

\maketitle

\section{Introduction}
\label{sec.intro}

\begin{figure}[!b]
\centering
\includegraphics[width=\linewidth]{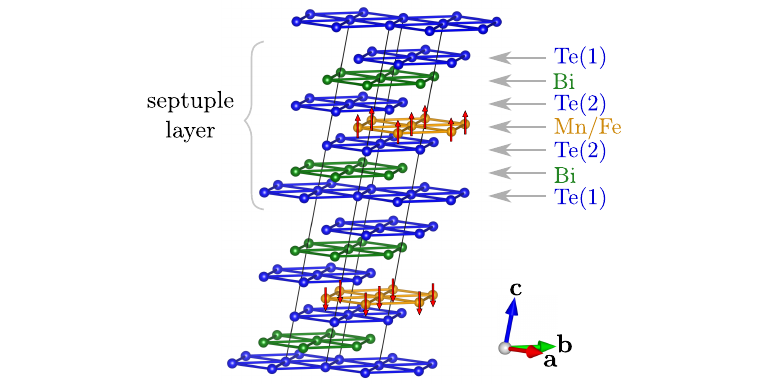}
\caption{
Magnetic unit cell of antiferromagnetic topological insulator $T$Bi$_{2}$Te$_{4}$ ($T=$Mn, Fe).
\label{fig.schemat}
}
\end{figure}

MnBi$_{2}$Te$_{4}$ is the first intrinsic antiferromagnetic (AFM) topological insulator~\cite{otrokov.klimovskikh.19,gong.guo.19} that has been recently observed experimentally.
Similarly to the other topological insulators (TI), this system possesses a layered structure (Fig.~\ref{fig.schemat}), while layers are bonded by the van der Waals interaction.
The AFM order exist below $T_\text{N} = 25$~K~\cite{yan.zhang.19,ding.hu.20,li.liu.20}, and is formed within Mn sublayers.
Additionally, an interplay between the topological properties of this system and the intrinsic magnetic order allows for the realization of many quantum phenomena including quantum anomalous Hall effect~\cite{deng.yu.20,ovchinnikov.huang.20} or axion insulator state~\cite{liu.wang.20}.

The topological properties of the MnBi$_{2}$Te$_{4}$ are exhibited by the realization of the electronic surface states~\cite{he.liu.19,chen.fei.19,hao.liu.19,chen.xu.19,ko.kolmer.20,nevola.li.20}, which in similarity to the case of TI (Bi$_{2}$Se$_{3}$ or Bi$_{2}$Te$_{3}$) form Dirac cones.
However, contrary to the ordinary TI, due to the intrinsic magnetic order, the time reversal symmetry breaking occurs and the Dirac gap is observed in the surface states~\cite{shikin.estyunin.20,shikin.estyunin.21,ma.zhao.21}.
Nevertheless, these surface states can give a dominant contribution to the electron--phonon interaction~\cite{sobota.yang.14,heid.sklyadneva.17,benedek.mirerartes.20}.

More recently, a successful single crystal growth of FeBi$_{2}$Te$_{4}$ was reported~\cite{saxena.rani.20}. 
The structure of this compound and its physical characteristics are not yet confirmed in a theoretical study, but it seems to have topological properties similar to the MnBi$_{2}$Te$_{4}$.
In this paper, we present the theoretical study of the structure and lattice dynamics  performed for both compounds, MnBi$_{2}$Te$_{4}$ and FeBi$_{2}$Te$_{4}$.
We confirm the dynamical stability of both crystals and the consistency between calculated and measured data on both structural parameters and their Raman shifts.
Moreover, we calculate the circular polarization of Te and Bi atoms and demonstrate that the emergence of chiral modes in these materials is feasible.

The paper is organized as follows.
First, we briefly describe the computational details (Sec.~\ref{sec.calc}).
The phonon dispersions are presented and discussed in Sec.~\ref{sec.dyn}.
Next, in Sec.~\ref{sec.raman}, we discuss thoroughly the Raman active modes in relation to the available experimental and theoretical results.
In Sec.~\ref{sec.phchir}, we discuss the realization of the chiral phonons in the $T$Bi$_{2}$Te$_{4}$.
Finally, we conclude our findings in Sec.~\ref{sec.sum}.

\section{Calculation details}
\label{sec.calc}

The first-principles (DFT) calculations were performed using the projector augmented-wave (PAW) potentials~\cite{blochl.94} implemented in the Vienna Ab initio Simulation Package ({\sc vasp}) code~\cite{kresse.hafner.94,kresse.furthmuller.96,kresse.joubert.99}.
The calculations are made within the generalized gradient approximation (GGA) in the Perdew, Burke, and Ernzerhof (PBE) parametrization~\cite{pardew.burke.96}.
Strong local electron interaction on the $3d$ orbitals of transition metals were taken into account using DFT+U scheme~\cite{liechtenstein.anisimov.95}, with the intraorbital Coulomb parameter $U = 5.0$~eV, similarly to the earlier studies~\cite{hu.gordon.20}.
Additionally, we included the spin--orbit coupling (SOC) as well as the van der Waals (vdW) corrections within the Grimme scheme (DFT-D2)~\cite{grimme.06}.

The magnetic unit cells were optimized using $24 \times 24 \times 4$ {\bf k}--point $\Gamma$--centered grids in the Monkhorst--Pack scheme~\cite{monkhorst.pack.76}.
The energy cutoff for the plane-wave expansion is set to $450$~eV.
The condition for the breaking of the optimization loop was the energy difference of $10^{-6}$~eV and $10^{-8}$~eV for ionic and electronic degrees of freedom for subsequent steps.
The crystal symmetry was analysed using {\sc FindSym}~\cite{stokes.hatch.05} and {\sc SpgLib}~\cite{togo.tanaka.18}, while the momentum space analysis was done using {\sc SeeK-path} tools~\cite{hinuma.pizzi.17}.

The dynamical properties were calculated using the direct {\it Parlinski--Li--Kawazoe} method~\cite{parlinski.li.97}. 
Under this calculation, the interatomic force constants (IFC) are found from the forces acting on atoms when an individual atom is displaced.
The forces were obtained by the first-principle calculations with {\sc vasp} using 
the supercell containing $3 \times 3 \times 1$ magnetic unit cells and reduced $\Gamma$-centered $3 \times 3 \times 2$ ${\bm k}$ mesh.
The phonon dispersion and polarization vectors analyses were performed using the {\sc Alamode} software~\cite{tadano.gohda.14}.
The mode symmetries at the $\Gamma$ point were found by the {\sc Phonon} software~\cite{phonon}.

\begin{figure}[!b]
\centering
\includegraphics[width=\linewidth]{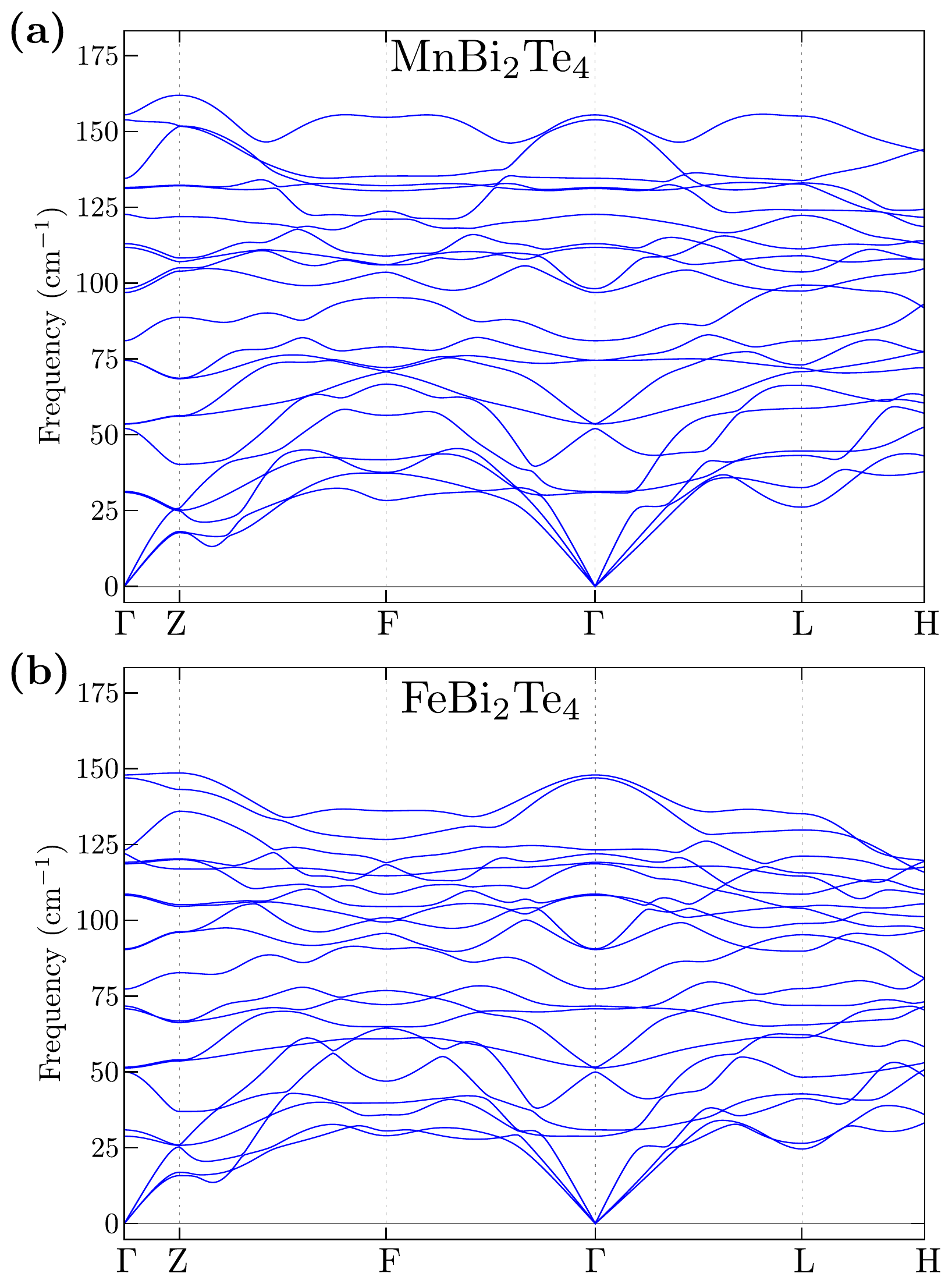}
\caption{
The phonon dispersion along high symmetry points.
\label{fig.phon}
}
\end{figure}

\section{Lattice dynamics}
\label{sec.dyn}

\paragraph*{Crystal structure. ---}
$T$Bi$_{2}$Te$_{4}$ crystallizes in the $R\bar{3}m$ (space group 166) rhombohedral structure presented in Fig.~\ref{fig.schemat}.
This system is composed of septuple layer (SL) slabs (containing the sequence of Te(1)--Bi--Te(2)--(Mn/Fe)--Te(2)--Bi--Te(1) atoms), separated by the vdW gap between Te(1) double-layer.
After the optimization, we obtain the lattice constants $a = 4.29$~\AA, and $c = 41.67$~\AA\ for MnBi$_{2}$Te$_{4}$, and $a = 4.26$~\AA, and $c = 41.46$~\AA\ for FeBi$_{2}$Te$_{4}$.
Our theoretical findings are in the excellent agreement with experimental results, i.e., $a = 4.33$~\AA, and $c = 40.92$~\AA\ for MnBi$_{2}$Te$_{4}$~\cite{aliev.amiraslanov.19} and $a = 4.39$~\AA, and $c = 42.69$~\AA\ for FeBi$_{2}$Te$_{4}$~\cite{saxena.rani.20}.
Atoms are located in Wyckoff position $6c$ ($0$,$0$,$z_{Bi}$), $6c$ ($0$,$0$,$z_{Te(1)}$), $6c$ ($0$,$0$,$z_{Te(2)}$), and $3c$ ($0$,$0$,$0$), for Bi, Te(1), Te(2), and $T$=Mn, Fe atoms, respectively.
We find $z_\text{Bi} = 0.42$, $z_\text{Te(1)} = 0.13$, and $z_\text{Te(2)} = 0.29$ for both compounds.

\begin{figure}[!b]
\centering
\includegraphics[width=\linewidth]{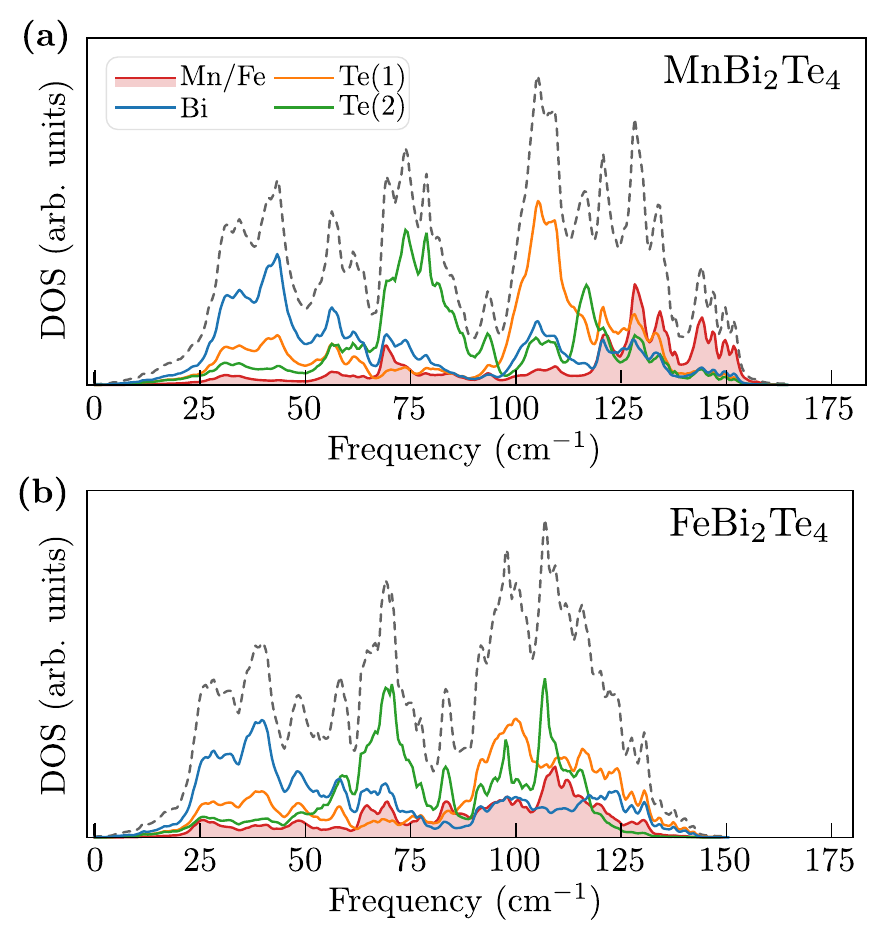}
\caption{
The total (dashed line) and partial (solid lines) phonon density of states.
\label{fig.phdos}
}
\end{figure}

\paragraph*{Phonon dispersion. ---}
The phonon dispersions of $T$Bi$_{2}$Se$_{4}$ are presented in Fig.~\ref{fig.phon}.
Introduction of the spin--orbit interaction in the calculations does not change the phonon dispersion qualitatively.
In the phonon dispersion, all acoustic branches show a linear dispersion in the vicinity of the $\Gamma$ point.
No imaginary phonon frequency was found for any system, which indicates the stability of these systems in the $R\bar{3}m$ phase.
For this reason, we can expect that the small substitution of Fe should be possible for the base system MnBi$_{2}$Se$_{4}$, contrary to the hole doping by Ca or Mg, which leads to substantial instability in the recently studied systems~\cite{han.sun.21}.

\paragraph*{Phonon density of states. ---}
The total and partial density of states are presented in Fig.~\ref{fig.phdos}.
As we can see, the vibrational modes of heavy Bi contribute mainly to the low range of frequencies.
Vibrations of Te in the first and second position [i.e. Te(1) and Te(2)], are placed in two separated frequency areas at the middle frequency region between $T=$(Mn,Fe) and Bi atomic layers.
The Te(1) atoms, which create a two-layer-like structure (see Fig.~\ref{fig.schemat}), vibrate with  higher frequencies (around $100$~cm$^{-1}$) than the Te(2) atoms that oscillate with average frequency of $75$~cm$^{-1}$.
Finally, $T=$(Mn,Fe) modes are located in the range of high frequencies. 
Interestingly, for both systems, the phonon dispersion associated with these modes creates a flat band around $130$~cm$^{-1}$ or $120$~cm$^{-1}$ in Mn or Fe compounds, respectively (see Fig.~\ref{fig.phon}).

\begin{table}[!t]
\caption{
\label{tab.activity}
Characteristic frequencies in cm$^{-1}$, activities (\mbox{R -- Raman}, \mbox{IR -- infrared}) and symmetries of the modes at $\Gamma$ point.
}
\begin{ruledtabular}
\begin{tabular}{rcc|rcc}
\multicolumn{3}{c}{MnBi$_{2}$Te$_{4}$} & \multicolumn{3}{c}{FeBi$_{2}$Te$_{4}$} \\ 
freq. & activity & symm. & freq. & activity & symm. \\
\hline 
30.65 & R & $E_\text{g}$ & 29.35 & R & $E_\text{g}$ \\
51.20 & R & $A_\text{1g}$ & 49.17 & R & $A_\text{1g}$ \\
52.67 & IR & $E_\text{u}$ & 50.60 & IR & $E_\text{u}$ \\ 
73.32 & R & $E_\text{g}$ & 70.05 & R & $E_\text{g}$ \\ 
76.66 & IR & $A_\text{2u}$ & 76.02 & IR & $A_\text{2u}$ \\
95.90 & IR & $E_\text{u}$ & 88.93 & IR & $E_\text{u}$ \\ 
110.51 & R & $E_\text{g}$ & 106.57 & R & $E_\text{g}$ \\
120.58 & R & $A_\text{1g}$ & 116.92 & IR & $E_\text{u}$ \\ 
129.16 & IR & $E_\text{u}$ & 119.82 & R & $A_\text{1g}$ \\
132.23 & IR & $A_\text{2u}$ & 121.05 & IR & $A_\text{2u}$ \\
151.21 & R & $A_\text{1g}$ & 144.23 & IR & $A_\text{2u}$ \\
152.84 & IR & $A_\text{2u}$ & 145.37 & R & $A_\text{1g}$
\end{tabular}
\end{ruledtabular}
\end{table}

\section{Raman active modes}
\label{sec.raman}

\paragraph*{Irreducible representations. ---}
The phonon modes of $T$Bi$_{2}$Te$_{4}$ at $\Gamma$ point can be decomposed into the irreducible representations of the space group $R\bar{3}m$ as follows:
\begin{eqnarray}
\nonumber \Gamma_\text{acoustic} &=& A_\text{2u} + E_\text{u} , \\ \nonumber \Gamma_\text{optic} &=& 3 A_\text{1g} + 3 A_\text{2u} + 3 E_\text{u} + 3 E_\text{g} .
\end{eqnarray}
In total, there are $21$ vibrational modes, seven nondegenerate $A_\text{1g}$ and $A_\text{2u}$ modes, and seven doubly degenerate $E_\text{u}$ and $E_\text{g}$ modes.
Here, optical vibrations $3 A_\text{2u} + 3 E_\text{u}$ are infrared active (IR), while optical modes $3 A_\text{1g} + 3 E_\text{g}$ are Raman (R) active.
The activity and symmetries of the modes at $\Gamma$ point for $T$Bi$_{2}$Te$_{4}$ are presented in Tab~\ref{tab.activity}.

The atoms participating in the vibrations of the $E_\text{g}$ and $E_\text{u}$ types oscillate in the SL plane ($a$--$b$ plane in Fig.~\ref{fig.schemat}).
Contrary to this, the modes $A_\text{1g}$ and $A_\text{2u}$ are related to out-of-plane oscillations (perpendicular to $a$--$b$ plane). 
Interestingly, the magnetic $T$ atoms do not participate in the vibrations of the Raman active modes (i.e. $A_\text{1g}$ and $E_\text{g}$).

\paragraph*{Selection rules for Raman-active modes. ---}
The non-resonant Raman scattering intensity depends in general on the directions of the incident and scattered light relative to the principal axes of the crystal. 
It is expressed by Raman tensor $R$, relevant for a given crystal symmetry, as~\cite{loudon.01}:
\begin{eqnarray}
\label{eq.intens} I \propto | e_{i} \cdot R \cdot e_{s} |^{2} ,
\end{eqnarray}
where $e_{i}$ and $e_{s}$ are the polarization vectors of the incident and scattered light. respectively.
According to the group theory, the Raman tensor for the $R\bar{3}m$ space group takes the following forms for the $A_\text{1g}$ and $E_\text{g}$ modes:
\begin{eqnarray}
\label{eq.raman} &R \left( A_\text{1g} \right) = & \left( \begin{array}{ccc}
a & 0 & 0 \\ 
0 & a & 0 \\ 
0 & 0 & b
\end{array} \right) \quad \text{and} \\
\nonumber & R \left( E_\text{g}^\text{I} \right) =& \left( \begin{array}{ccc}
c & 0 & 0 \\ 
0 & -c & d \\ 
0 & d & 0
\end{array} \right); \; R \left( E_\text{g}^\text{II} \right) =
\left( \begin{array}{ccc}
0 & -c & -d \\ 
-c & 0 & 0 \\ 
-d & 0 & 0
\end{array} \right) . 
\end{eqnarray}
In the backscattering configuration, $e_{i}$ and $e_{s}$ are placed within the $xy$ plane.
The polarization vectors for linearly polarized light in the $x$ and $y$ directions are \mbox{$e_{x} = \left( 1 \; 0 \; 0 \right)$} and \mbox{$e_{y} = \left( 0 \; 1 \; 0 \right)$}, respectively.
Similarly, the polarization vector for left $\sigma^{+}$ and right $\sigma^{-}$ circularly polarized light are \mbox{$\sigma^{\pm} = \frac{1}{\sqrt{2}} \left( 1 \; \pm i \; 0 \right)$}.
Using Eq.~(\ref{eq.intens}) and the Raman tensors~(\ref{eq.raman}), we can determine the selection rules and Raman intensities for various scattering geometries.
Tab.~\ref{tab.ramanselect} summarized the Raman response in the backscattering geometry for four polarization configurations.
As we can see, it is possible to distinguish the $A_\text{1g}$ and $E_\text{g}$ using the different backscattering configurations.
The same holds for both linear and circularly polarized Raman spectra measurements.
For example, this property of the Raman modes was used  to differentiate between Raman active modes using co- and cross-circular Raman back-scattering in Ref.~\cite{cho.kang.21}.
The group theory study of the vibrational modes in bulk paramagnetic $T$Bi$_{2}$Te$_{4}$, the phonon selection rules, and the real-space displacements corresponding to each mode are presented in Ref.~\cite{rodriguezvega.leonardo.20}.

\begin{table}[!t]
\caption{
\label{tab.ramanselect}
Selection rules for Raman-active modes.
}
\begin{ruledtabular}
\begin{tabular}{lcc}
configuration & $A_\text{1g}$ & $E_\text{g}$ \\
\hline 
$e_{x}$ in $e_{x}$ out (linear $\parallel$) & $|a|^{2}$ & $|c|^{2}$ \\
$e_{x}$ in $e_{y}$ out (linear $\perp$) & $0$ & $|c|^{2}$ \\
$\sigma^{+}$ in $\sigma^{+}$ out (cocircular) & $2|a|^{2}$ & $0$ \\
$\sigma^{+}$ in $\sigma^{-}$ out (cross-circular) & $0$ & $2|c|^{2}$
\end{tabular}
\end{ruledtabular}
\end{table}

\paragraph*{Comparison with related results. ---}
In Tab.~\ref{tab.ramancomp}, we present a comparison of our theoretically obtained frequencies of the Raman active modes with previous results (both experimental and theoretical).
It should be mentioned that not all Raman modes can be observed at each measurement.
This is caused by the relatively weak intensity of some Raman modes and the insensitivity of the used method for observing low-frequency modes.
For example, for MnBi$_{2}$Te$_{4}$, in Ref.~\cite{aliev.amiraslanov.19} or~\cite{pei.xia.20} only four Raman frequencies are observed, while in Ref.~\cite{cho.kang.21} we can distinguish five distinctive peaks.
Similarly, in the case of the FeBi$_{2}$Te$_{4}$, the Raman spectroscopy experiment exhibits three distinct phonon modes at $65$~cm$^{-1}$, $110$~cm$^{-1}$, and $132$ cm$^{-1}$ along with two split secondary modes at $90$~cm$^{-1}$, and $144$~cm$^{-1}$~\cite{saxena.rani.20}.
The measured frequencies are slightly lower than the values obtained in calculations. 
The biggest difference of $\sim 12$~cm$^{-1}$ is observed for the high-frequency A$_{1g}$ mode.
The reason for that can be the temperature at which the measurement is carried out. 
Recent experiments on MnBi$_{2}$Te$_{4}$ show that the increase of temperature~\cite{cho.kang.21} leads to the shift of the Raman frequency modes to lower values.
Similar temperature dependence of the Raman frequencies was  observed for isostructural  materials, PbBi$_{2}$Te$_{4}$~\cite{mal.bera.19} and GeBi$_{2}$Te$_{4}$~\cite{singh.rawat.22}.
In contrast, the external pressure imposed on the  MnBi$_{2}$Te$_{4}$ crystal causes the increase of mode frequencies~\cite{pei.xia.20}.

\begin{table}[!t]
\caption{
\label{tab.ramancomp}
Comparison of the Raman active modes frequencies (cm$^{-1}$) for different compounds.
}
\begin{ruledtabular}
\begin{tabular}{llllllll}
$E_\text{g}$ & $A_\text{1g}$ & $E_\text{g}$ & $E_\text{g}$ & $A_\text{1g}$ & $A_\text{1g}$ &  &\\
\hline 
\multicolumn{8}{c}{\bf {MnBi$_{2}$Te$_{4}$}} \\
\hline
 30.65 & 51.20 & 73.32 & 110.51 & 120.58 & 151.21 & {\bf this work} & \\ 
 --- & 46 & 65 & 102 & --- & 138 & measurement & \cite{aliev.amiraslanov.19} \\ %OK
 --- & 47.4 & 67.4 & 104.2 & --- & 139.8 & measurement & \cite{pei.xia.20} \\ %OK
  --- & 45.7 & 66.9 & 103.4 & 113.5 & 138.6 & measurement & \cite{cho.kang.21} \\ %OK
 26.9 & 48.0 & 68.3 & 105.1 & 115.6 & 140.0 & measurement & \cite{choe.lujan.21} \\ %OK
 --- & 47 & 60 & 105 & --- & 142 & calculation & \cite{aliev.amiraslanov.19} \\ %OK
 30.3 & 46.0 & 74.5 & 109.6 & 117.4 & 144.1 & calculation & \cite{cho.kang.21} \\ %OK
 29.4 & 51.8 & 73.4 & 109.1 & 119.4 & 149.1 & calculation & \cite{choe.lujan.21}\\ %OK
\hline 
\multicolumn{8}{c}{\bf {FeBi$_{2}$Te$_{4}$}} \\
\hline 
 29.35 & 49.17 & 70.05 & 106.57 & 119.82 & 145.37 & {\bf this work} & \\
 --- & --- & 65 & --- & 110 & 132 & measurement & \cite{saxena.rani.20}\\ %OK
\hline 
\multicolumn{8}{c}{\bf{Bi$_{2}$Te$_{3}$}} \\
\hline 
 --- & 62.5 & --- & 103 & --- & 134 & measurement  & \cite{richter.becker.77} \\ %OK
 --- & 61.5 & --- & 101.5 & --- & 133.5 & measurement  & \cite{chris.sklyadneva.12} \\ %OK
 --- & 62.3 & --- & 103 & --- & 134 & measurement  & \cite{goncalves.couto.10} \\ %OK
 34.4 & 62.1 & --- & 101.7 & --- & 134.0 & measurement  & \cite{shahil.hossain.10} \\ %OK
 42.4 & 66.4 & --- & 106.7 & --- & 134.8 & calculation & \cite{wang.zhang.12} \\ %OK
 42.1 & 64.2 & --- & 112.3 & --- & 139.2 & calculation & \cite{chris.sklyadneva.12} \\ %OK
 50.6 & 71.1 & --- & 118.5 & --- & 128.3 & calculation & \cite{jenkins.rayne.72} %OK
\end{tabular}
\end{ruledtabular}
\end{table}

Finally, we briefly describe the differences between the Raman frequencies in  $T$Bi$_{2}$Te$_{4}$ and in the parent Bi$_{2}$Te$_{3}$ material.
In the studied materials, the $T$Te layer is inserted in van der Waals gaps of Bi$_{2}$Te$_{3}$ layers. 
Due to similar crystal structures described by the same space group symmetry, the four Raman modes of Bi$_{2}$Te$_{3}$ correspond to four Raman modes of $T$Bi$_{2}$Te$_{4}$. 
The smaller number of Raman active modes is caused by the smaller number of atoms in the unit cell. 
The modifications of the crystal structure induced by additional $T$Te layers lead to the changes of distance and strength of the covalent bonds between pairs of atoms.
This, in turn, results in the shift of Raman modes (cf. Tab.~\ref{tab.ramancomp}).

For example, the frequency of the first $A_\text{1g}$ mode is reduced from $\sim62$~cm$^{-1}$ in Bi$_{2}$Se$_{3}$ to $\sim50$~cm$^{-1}$ in $T$Bi$_{2}$Se$_{4}$, while the high-frequency $A_\text{1g}$ is shifted from $\sim 135$~cm$^{-1}$ for Bi$_{2}$Te$_{3}$ to $\sim 140$~cm$^{-1}$ for $T$Bi$_{2}$Te$_{4}$.
Contrary to this, the $E_\text{g}$ modes are almost unchanged.
As we mentioned previously, $A_\text{1g}$ and $E_\text{g}$ modes realize the out-of-plane and in-plane vibrations, respectively~\cite{rodriguezvega.leonardo.20}, and thus the distance between Te--Bi plays a crucial role~\cite{mal.bera.19}.
Because the covalent interactions are much stronger along Te(1)--Bi--Te(2) than between Te--Te and Bi--Bi pairs, only the $A_\text{1g}$ mode is shifted.

% stale silowe PRB 88 035402

\begin{figure*}
\centering
\includegraphics[width=\linewidth]{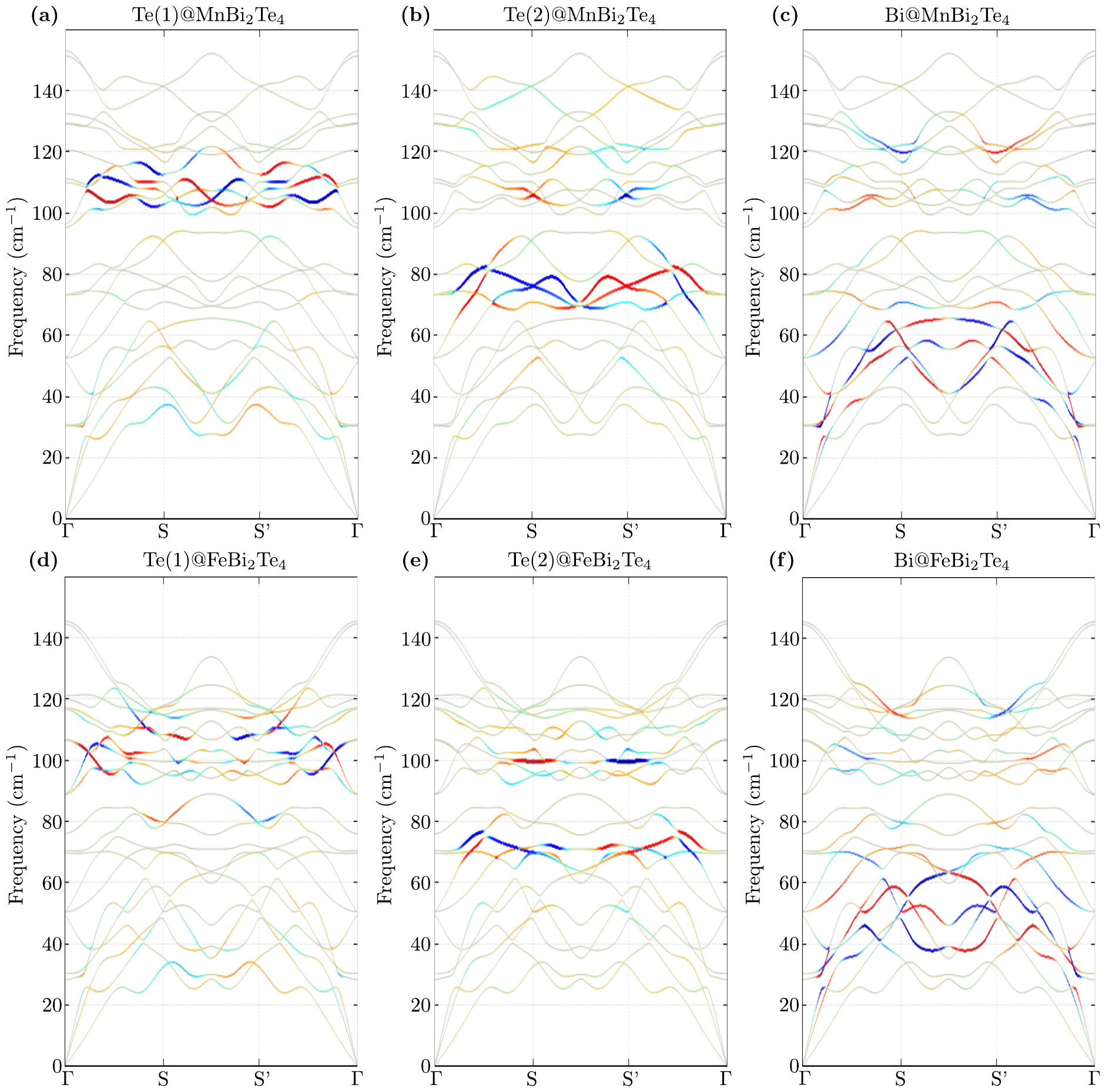}
\caption{
Circular polarization of Te(1), Te(2), and Bi (panels from left to right) for MnBi$_{2}$Te$_{4}$ and FeBi$_{2}$Te$_{4}$ (top and bottom panels, respectively).
Red and blue lines denotes left and right handed circular polarization of phonons, while the line width corresponds to the circulation value.
\label{fig.phchir}
}
\end{figure*}

\section{Circular phonon polarization}
\label{sec.phchir}

The helicity of incident photons is completely reversed in the Raman process involving the double degenerated $E_\text{g}$ modes.
This phenomenon is highly relevant in the context of circularly polarized phonons~\cite{zhang.niu.14,liu.lian.17,zhang.niu.15,chen.wu.19,ptok.kobialka.21}.
Indeed, $T$Bi$_{2}$Te$_{4}$ possesses the $C_{3v}$ symmetry (each layer of atoms inside the SL makes  a triangular lattice), which allows for the realization of circularly polarized phonons~\cite{coh.19}.
Although none of modes at $\Gamma$ point have the intrinsic chirality, we can obtain the chiral phonons via superimposing the doubly degenerate $E_\text{g}$ modes~\cite{chen.zhang.18}.
Nevertheless, the chiral phonons can emerge out of the center of the Brillouina zone (i.e. away from the $\Gamma$ point), where degeneracy is lifted.

The circular polarization of the phonons can be studied by analyzing the phonon polarization vector, which can be found from the diagonalization of the dynamical matrix:
\begin{eqnarray}
\nonumber D_{\alpha\beta}^{jj'} ( {\bm q} ) \equiv \frac{1}{\sqrt{m_{j}m_{j'}}} \sum_{n} \Phi_{\alpha\beta} ( j0, j'n ) \exp \left( i {\bm q} \cdot {\bm R}_{j'n} \right) , \\
\end{eqnarray}
where ${\bm q}$ is the phonon wave vector and $m_{j}$ denotes the mass of $j$th atom. 
Here $\Phi_{\alpha\beta} (j0,j'n)$ is the IFC tensor ($\alpha$ and $\beta$ denotes the direction index, i.e. $x$, $y$, and $z$) between $j$th and $j'$th atoms located in the initial ($0$) and $n$th primitive unit cell.
Then, the phonon spectrum as well as the polarization vectors are given by the eigenproblem of the dynamical matrix:
\begin{eqnarray}
\omega_{\varepsilon{\bm q}}^{2} \text{e}_{\varepsilon{\bm q}\alpha j} = \sum_{j'\beta} D_{\alpha\beta}^{jj'} \left( {\bm q} \right) \text{e}_{\varepsilon{\bm q}\beta j'} .
\end{eqnarray}
Here, the $\varepsilon$ branch describes the phonon with a frequency $\omega_{\varepsilon{\bm q}}$ and a polarization vector $\text{e}_{\varepsilon{\bm q}\alpha j}$. 
Each $\alpha j$ component of the polarization vector is associated with the displacement of the $j$th atom in the $\alpha$th direction.

The phonon mode related to the particular polarization vector $\text{e}_{\varepsilon{\bm q}\alpha j}$ can be discussed in the context of the circular polarization.
For this purpose, we introduce a new basis defined as~\cite{zhang.niu.15,chen.wu.19,ptok.kobialka.21}:
\mbox{$\vert R_{1} \rangle \equiv \frac{1}{\sqrt{2}} \left( 1 \; i \; 0 \cdots 0 \right)^{T}$}; 
\mbox{$\vert L_{1} \rangle \equiv \frac{1}{\sqrt{2}} \left( 1 \; -i \; 0 \cdots 0 \right)^{T}$};
\mbox{$\vert Z_{1} \rangle \equiv \frac{1}{\sqrt{2}} \left( 0 \; 0 \; 1 \cdots 0 \right)^{T}$};
$\cdots$; 
\mbox{$\vert R_{j} \rangle \equiv \frac{1}{\sqrt{2}} \left( \cdots 1 \; i \; 0 \cdots 0 \right)^{T}$}; 
\mbox{$\vert L_{j} \rangle \equiv \frac{1}{\sqrt{2}} \left( \cdots 1 \; -i \; 0 \cdots 0 \right)^{T}$};
\mbox{$\vert Z_{j} \rangle \equiv \frac{1}{\sqrt{2}} \left( \cdots 0 \; 0 \; 1 \cdots 0 \right)^{T}$};
$\cdots$. 
It means that two in-plane components are replaced by the circular polarization vectors $\sigma^{\pm}$ (defined in Sec.~\ref{sec.raman}), while the third component is unchanged.
In this basis, each polarization vector, $\text{e} \equiv \text{e}_{\varepsilon{\bm q}\alpha j}$, is represented as:
\begin{eqnarray}
\text{e} = \sum_{j} \left( \alpha_{j}^{R} \vert R_{j} \rangle + \alpha_{j}^{L} \vert L_{j} \rangle + \alpha_{j}^{Z} \vert Z_{j} \rangle \right) ,
\end{eqnarray}
where $\alpha_{j}^{V} = \langle V_{j} \vert \text{e} \rangle$, for $V \in \{ R , L , Z \}$  and $j \in \{ 1 , 2 , \cdots , N \}$ (N is a total number of atoms in a primitive unit cell).
The way of the $j$th atom movement is determined by a circulation
$\mathcal{C} = | \alpha_{j}^{R} |^{2} - | \alpha_{j}^{L} |^{2}$.
When $\mathcal{C} = 0$, the atom is involved in an ordinary non-circular vibration, 
and when $\mathcal{C} \neq 0$, it realizes a circular motion. 

In $T$Bi$_{2}$Te$_{4}$, $T$ atoms do not exhibit circular vibrations, being the only magnetic atoms in the compound. 
The other atoms, Te(1), Te(2), and Bi, realize a circular motion with different frequencies.
Fig.~\ref{fig.phchir} presents a circular polarization of Te(1), Te(2), and Bi, along a path between two $\Gamma$ points in $a$--$b$ plane (cf. Fig.~\ref{fig.schemat}).
As the system possesses the inversion symmetry, atoms from opposite sites of the SL layer have an opposite circulation (i.e. the total circulation of the system is zero).
The chiral phonons of Te(1), Te(2), and Bi are located around $110$~cm$^{-1}$, $75$~cm$^{-1}$, and $50$~cm$^{-1}$.
This ``separation'' of different types of phonons generating a circular motion of atoms is in agreement with the discussion of the phonon DOS (cf.~Sec.~\ref{sec.dyn}).

\section{Summary}
\label{sec.sum}

In this paper, we investigated the dynamical properties of $T$Bi$_{2}$Te$_{4}$ ($T=$Mn, Fe) compounds.
Both systems crystallize in the $R\bar{3}m$ rhombohedral structure.
The phonon dispersion relations do not exhibit any imaginary frequencies and thus, both compounds are dynamically stable.
We performed the analyses of the mode symmetries at the $\Gamma$ point.
The Raman and infrared active modes were calculated.
The frequencies of the Raman active modes are in agreement with previous theoretical and experimental results.
We point out that the Raman backscattering using co- or cross-circular configuration can be useful in distinguishing the $A_\text{1g}$ and $E_\text{g}$ Raman active mode.
Additionally, we show that the initially double degenerate $E_\text{g}$ modes (at the $\Gamma$ point), give rise to emergence of the circularly polarized modes away from the $\Gamma$ point.
Moreover, chiral phonon modes are separated in the frequency domain depending on the atomic layer within the septuple layer.

\begin{acknowledgments}
Some figures in this work were rendered using {\sc Vesta}~\cite{momma.izumi.11}.
This work was supported by National Science Centre (NCN, Poland) under Projects No.
2018/31/N/ST3/01746 (A.K.),
2017/25/B/ST3/02586 (M.S.),
and
2016/21/D/ST3/03385 (A.P.).
In addition, A.P. appreciates funding in the frame of scholarship of the Minister of Science and Higher Education of Poland for outstanding young scientists (2019 edition, No.~818/STYP/14/2019).
\end{acknowledgments}

\newpage

%\nocite{*}
\bibliography{biblio}

\end{document}